\documentclass[
aps,nofootinbib,
showpacs,showkeys,preprint
tightenlines,preprintnumbers,] {revtex4}

\usepackage{epsf,epsfig,subfigure,
graphicx,amsmath,amssymb}
\usepackage{color}

\newcommand{\dis}[1]{\begin{equation}\begin{split}#1\end{split}\end{equation}}

\newcommand{\gev}{\,\textrm{GeV}}
\newcommand{\mev}{\,\textrm{MeV}}

\newcommand{\LQCD}{\Lambda_{\rm QCD}}

 \def\thb{\bar{\theta\,}}

\def\eVV{\,\textrm{eV}}

\def\etal{{\it et al}}
\def\ie{{\it i.e.~}}

\def\EE8{E$_8\times$E$_8^\prime$}

\def\qg{q\&g}
   
\begin{document}

\title{``Invisible'' QCD axion rolling through the QCD phase transition 
} 
   
\author{Jihn E.  Kim$^{1,2}$, Se-Jin  Kim$^{2}$ }
\address
{ 
$^{1}$Center for Axion and Precision Physics Research (Institute of Basic Science), KAIST Munji Campus, Munjiro 193,  Daejeon 34051, Republic of Korea,  \\ 
$^{2}$Department of Physics, Kyung Hee University, Seoul 02447, Republic of Korea 
}

\begin{abstract} 
Visible matter in the current Universe is a consequence of the phase transition of the strong force, quantum chromodynamics (QCD). This phase transition   has occurred at the Universe temperature around $T_c\simeq 165\,$MeV while it was expanding. Strongly interacting matter particles are quarks above $T_c$, while they are pions, protons and neutrons below $T_c$. The spin degrees of freedom 37 ($u$ and $d$ quarks and gluons)   just above $T_c$ are converted to 3 (pions) after the phase transition.  This phase transition might have been achieved mostly at  supercooled temperatures. The supercooling was provided by the expansion of the Universe. We obtain  the effective bubble formation rate $\alpha(T)\approx 10^{4-5}\,$MeV  and  the completion temperature of the phase change (to the hadronic phase),  $T_f\simeq 126\,$MeV.  During the phase transition,  the scale factor $R$ has increased by a factor of 2.4. This provides a key knowledge on the energy density of ``invisible''  QCD axion at  the full hadronic-phase commencement temperature $T_f$, and allows for us to estimate the current energy density of cold dark matter composed of  ``invisible''  QCD axions.

\end{abstract}
\maketitle


\section{Introduction and summary}\label{sec:Introduction}

 An ``invisible'' QCD axion \cite{KSVZ1,KSVZ2,DFSZ,DFSZ2} attracted a great deal of attention   because of its solution to the strong CP problem \cite{PQ77} and its  role as cold dark matter (CDM) candidate in the evolving Universe \cite{Preskill83,Abbott83,Dine83}. The invisible axion  is a descendent of the Peccei-Quinn (PQ) symmetry, while the earlier electroweak scale axion \cite{Weinberg78,Wilczek78} is not relevant for the present study which relates the cosmic axion density to the detection possibility.  For the strong CP solution, only one axion is needed, which is the phase of a complex singlet field \cite{KSVZ1} in the Standard Model.  Recently it attracted a great deal of attention in the lattice community, where it became possible to accomodate the effects of fermion loops now \cite{Boyarski16}.  
 
The ``invisible'' QCD axion $a$ (or $\bar\theta=a/f_a)$  has arisen as a favored CDM candidate based on the reasonable estimates on the energy of bosonic collective motion (BCM) in the Universe   \cite{KimSemTsu14}.    An important parameter for this determination is the time $t_1$ when the temperature dependent axion mass equals the Hubble parameter,  $m_a=3H$. The usual axion window for CDM assumes the misalignment angle $\bar\theta_1=$O(1) at cosmic time  $t_1$ corresponding to the Universe temperature $\simeq 1\,\gev$, which decreases to the current value by a factor $\approx 10^{-19}$ for the case of ``invisible'' axion \cite{Sikivie17}.\footnote{Based on this number, the Rochester-Brookhaven and Univ. of Florida groups started to detect ``invisible'' axions \cite{KimRMP10}. The recent ADMX report reached the line roughly a factor 2 above the KSVZ line at $m_a\simeq 2.4\times 10^{-5}\eVV$ \cite{ADMX18}.} In this estimation, there are five important facts to be stressed. Firstly, the effects of anharmonic terms are important for large values of $\thb$ \cite{Turner86,Bae08}. If the initial $\thb_1$ is close to $\pi$, then there was a long period of time when $\thb$ did not roll, \ie the presence of bottle-neck period. Second, after the QCD phase transition, $\thb$ undergoes a   decrease for which the diminishing factor is now reliably given  in Ref. \cite{KimKimNam18}. Third,  the QCD phase transition undergoes under the thermodynamic principles, asking for the basic equation for the QCD phase transition in the evolving Universe. The fourth is obtaining the finishing time $t_f$ of this QCD phase transition.  The fifth is relating these to the axion energy density after the QCD phase transition.  Here we discuss   the last three questions.  Another new feature here is that  different $\thb_1$'s are used for different axion masses, in contrast to one value of  $\thb_1$ for any axion mass  in the previous studies \cite{Turner86,Bae08}.
 
Related to the third question, the conservation of Gibb's free energy during the QCD phase transition is adopted.
 One can understand why it restricts the evolution so much just by counting the number of degrees in the quark and gluon($\qg$-)  and hadronic($h$-) phases above and below the critical temperature $T_c$, respectively. For up and down quarks and gluons, the number of degrees is  37 in the $\qg$-phase and 3 (pions) in the $h$-phase. In countiung 3 pions, we neglect the baryon number due to the small $\Delta B$ asymmetry of order $10^{-9}$.  To use  the  Gibbs free energy conservation, we will determine  the pion number density first in the $h$-phase.
At and below the critical temperature $T_c$, both $\qg$- and $h$-phases co-exist 
with the same Gibbs free energy, which is relevant because the two phases have the same temperature and pressure.  Next, at an appropriate super-cooled temperature,  the $h$-phase bubbles start to expand.  This condition determines the parameter $\alpha$ in our evolution equation of the fraction of  $h$-phase bubbles, Eq. (\ref{eq:diffequa}). Using this equation, we obtain the finishing time $t_f$. Then, we obtain the axion energy density at time $t_f$.

\begin{figure}[!t]
\includegraphics[width=13cm,height=9cm]{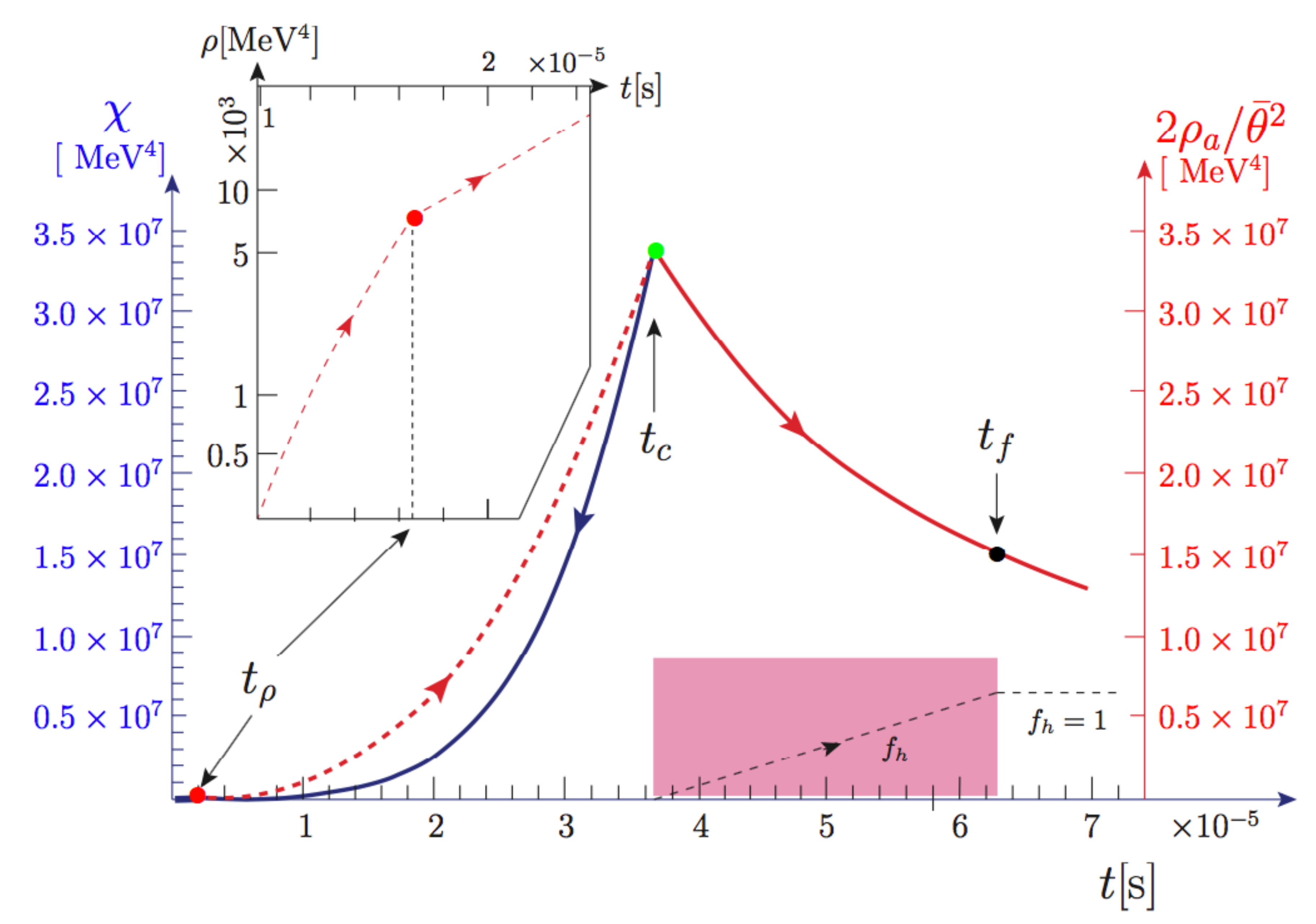}
 \caption{A view of susceptibility $\chi$ (the blue curve) and axion energy density $\rho_a$ (the red curve).  $\rho_a$ is in units of $\frac12\thb^2(t)$ near $\thb=0$ for $C=1.5$ from $t_c$ to the finishing point $t_f$ of the phase transition.  
 Below the $\rho$ meson scale, we used two temperature powers to connect to the $h$-phase value at $T_c$ smoothly. In the logarithmically enlarged inset, two different powers show a cusp. For $m_a(0)< 3.58\times 10^{-3}\eVV$, we have $t_1>t_\rho$. }\label{fig:rhoa}
\end{figure}
In Fig. 1, we show the susceptibility $\chi$ (the blue curve) and axion energy density $\rho_a$ (the red curve) for $C=1.5$ as functions of cosmic time $t$.   
 $C$  used in Fig. 1 is the parameter representing the effects of bubble coalition which appears in Eq. (\ref{eq:diffequa}). The critical point $t_c$ is special in that there one can calculate relativistic degrees of freedom without ambiguity both in the $\qg$- and $h$-phases. The only condition is that the Gibbs free energies are the same in both phases. When we consider this point as the beginning of the phase transition, we can take all states are in the $\qg$-phase. So, before the phase transition commence, we can go back to earlier cosmic time with 100\,\% $\qg$ degrees to estimate relativistic degrees above $T_c$. Susceptibility $\chi$ marked with the blue curve is calculated with this assumption.  For this calculation, only the temperature information due to instanton effects \cite{Pisarski81} is needed without any need for the cosmic evolution. This blue part is purely strong interaction effect. To connect to the $h$-phase value at $T_c$ smoothly, we used two temperature powers, which is manifested by the cusp in the inset. In the logarithmically enlarged inset, two different powers show a cusp. $m_a(0)= 10^{-4}\eVV$ is used for the curves in  Fig. 1. If  $m_a(0)< 3.58\times 10^{-3}\eVV$, we have $t_1>t_\rho$, which invalidates our study. Using this blue curve, we calculate $t_1$, and  Fig. 1 is for $m_a(0)= 10^{-4}\eVV$. For many different $m_a(0)$, we calculate different $t_1$'s. From this time $t_1$, we solve the axion field equation in the Universe to estimate the axion energy density, which are marked first by the dashed red  curve up to $t_c$ then as the solid red curve down to $t_f$. This evolution for different axion masses is presented in Fig. 5.

 The zero temperature expression of $\chi$, \ie in the $h$-phase, was given in \cite{BardeenTye78} and  the high temperature  expression was given in \cite{Weinberg78,Baluni78}.  The recent estimates of $\chi$ around the QCD phase transition have been performed from the  lattice calculation, including  the temperature effect  \cite{Boyarski16,ICTP16}. The lattice calculation must give the earlier zero temperature value \cite{BardeenTye78}. The earlier high temperature expression gave a temperature dependence   but its overall coefficient was not given \cite{Pisarski81}. We calculated this overall coefficient by the relativistic degrees given at $t_c$ as described above.   We used the powers $ T^{-8.16}$ \cite{Pisarski81} for $T>m_\rho\simeq 770\mev$  and $T^{-4.21}$  for $T_c< T<770\mev$. 
 An important aspect to be noted  is that  the QCD phase transition has occurred during the evolution of the Universe, as shown by the red curve in  Fig. 1.  If the Universe evolution does not allow a completion of this phase transition, the current Universe may look like a Swiss cheese and the homogeneous one has never arisen.  The fraction of $h$ phase, $f_h$, is shown in the lavender square. Time $t_f$ is the completion time of the QCD phase transition.

Multiplying all these factors, we obtain the current vacuum angle $\thb_{\rm now}$,
  \dis{
\thb_{\rm now}\simeq \bar{\theta}_1\cdot r_{f/1}\cdot \left(\frac{\thb_{\rm now}}{\thb_f}\right).
\label{eq:thbratio}
}
This expression shows how to estimate $\thb_{\rm now}$ from the initital misalignment angle $\bar{\theta}_1$ if we know two factors $r_{f/1}=r_{\rm osc/1}\cdot r_{f/\rm osc}$ and $ \thb_{\rm now}/\thb_f$. In this paper,  $r_{f/1}$ is calculated  and  $\thb_{\rm now}/\thb_f$ is estimated in  \cite{KimKimNam18}.   $\thb_{\rm now}$ is the important parameter, appearing in the expression of   the current CDM axion energy density.
 
\section{QCD phase transition} \label{sec:QCD}

In our study of QCD phase transition, it is sufficient to consider  up and down quarks, $u$ and $d$.\footnote{Addition of the strange quark $s$ would change parameters at a 5\,\%  level, viz. $m_d/m_s\simeq 1/20$.} 
The chiral symmetry breaking is proportional to light quark masses $m_um_d/(m_u+m_d)$  since it should vanish if any one quark is massless. Even if the QCD scale $\LQCD$ is a few hundred MeV, the axion mass should take into account the chiral symmetry breaking in terms of the current quark masses $m_u$ and $m_d$ in the $\qg$-phase.   In most literatures, axion energy density uses the parameters of the  $h$-phase, \ie $m_{\pi^0}$ and $f_{\pi^0}$. This expression should encode the chiral invariance in case of $m_u=0$ or $m_d=0$.   In the lattice calculation, topological susceptibility $\chi$ is calculated for $\langle \bar{q}_iq_i\rangle$ for quark fields $q_i$.  
Summarizing these for two flavors \cite{KimPRP87},\footnote{For $\thb$ near 0, the coefficient in the $\qg$ phase becomes  $({m_u}\LQCD/{1+Z})^2$.}
\dis{
&\textrm{Quark and gluon phase with~}\LQCD:~ f_a^2m_a^2= \frac{(\sin^2\thb/\thb^2)}{2Z\cos\thb+1+Z^2} \,m_u^2\LQCD^2\left(\frac12 \thb^2\right),\\[0.5em]
&\textrm{Hadronic phase in terms of~}f_{\pi^0}^2 m_{\pi^0}^2:~  f_a^2m_a^2= \frac{Z\,(\sin^2\thb/\thb^2)}{2Z\cos\thb+1+Z^2}\,f_{\pi^0}^2 m_{\pi^0}^2\,\left(\frac12 \thb^2\right) ,\\[0.5em]
&\textrm{Lattice susceptibility~}\chi :~ f_a^2m_a^2= \chi\,\left(\frac12 \thb^2\right) ,\label{eq:AxPhmasses}
}
where $Z=m_u/m_d$ and $\chi\simeq (76\,\mev)^4$ \cite{Boyarski16,ICTP16} and $\left(\frac12 \thb^2\right)$ is simply denoting  the axion field operator divided by $f_a^2$.  The $\qg$-phase expression is consistent with the symmetry expression ${\cal L}_{\bar\theta}\propto \bar{\theta}m_u\LQCD^3/(1+Z)$ in case $\thb\simeq 0$ \cite{KimRMP10}. 
The susceptibility presented in Eq. (\ref{eq:AxPhmasses})  gives the axion mass and also the effect of anharmonic terms. Of course, the vacuum is taken at $\thb=0$ but also large effects of the anharmonic terms are present when we take large values of $\thb$ in Eq. (\ref{eq:AxPhmasses}). The factor $(\frac12\thb^2)$ is not included in the definition of $\chi$ and is written just for a reference that its coefficient is the axion mass if devided by $f_a^2$ in the limit $\thb=0$.
The coefficient in the $\qg$-phase is negligible near the region $\thb=\pi$. It will be very difficult to discuss this region because of this singular behavior. Eexcept this singular region, we presented the susceptibility for $\thb=1$ if not explicitly stated. 

 If we use $m_u\simeq 2.5\,\mev, Z=1/2$ \cite{Manohar16}  and $\LQCD\approx \LQCD^{(3)}\simeq 332\,\mev$ \cite{QCD16}, the values in Eq. (\ref{eq:AxPhmasses}) for the $\qg$  and  the  $h$ phases are  $(88\,\mev)^4$ and  $(77\,\mev)^4$, respectively. Note that the values calculated in the $h$-phase, the second and third lines of (\ref{eq:AxPhmasses}), are almost identical. This confirms the validity of the lattice calculation in the $h$-phase \cite{Boyarski16,ICTP16,Sharma16}. 
  
Early works on the QCD phase transition in the lattice community were dominated by quenched results \cite{Kogut88}, and claimed the first order phase transition  as,  ``It is numerically well-established the phase transition  is the first order in the quenched limit, and there is strong numerical evidence for first order in the chiral limit" \cite{Forcrand07}. On the other hand,  the cross over transition was observed in Ref. \cite{AokiY06}. The recent developments in saving computing time, using M\"{o}bius parameters, confirmed the crossover phase transition \cite{LosAlamos14} because of the failure of growth of susceptibility ``$\chi$ as $2^3$ when the volume is increased from $32^3$ to $64^3$", and claimed ``the QCD phase transition is not first order but a cross-over."   The critical temperature was given as $154\pm1\pm8\,\mev$, where chiral quarks appear above 164\,MeV. The cross-over begins with the second order without growing $\chi$ and finishes as the first order in the end.  Since the hint of the cross over transition appears around 164\,MeV  \cite{LosAlamos14}, we use the critical temperature  $T_c$ given in \cite{ICTP16} 
\dis{
T_c=165\,\mev.\label{eq:TcLatt}
}

So, if $h$-phase bubbles form inside the $\qg$-phase,  formation of one typical bubble size will be dominated, which we take as $R_i^3$.
Here, we adopt two basic principles:  (i) two phases ($\qg$- and $h$-phases) coexist in the principle of conserved Gibbs free energy  \cite{HuangStat}, and (ii) pion bubbles start to expand at supercooled temperatures. The phase transition is completed into the $h$-phase by the time $t_f$ in the evolving Universe.  Contributions to energy density and entropy from light degrees below $\LQCD$, in the initial and final states, are 
\dis{
\textrm{Before}&\left\{\begin{array}{l}\rho  =\frac{\pi^2}{30}\,g_*^i T^4\\[0.5em]
s  =\frac{2\pi^2}{45}\,g_*^i T^3\\[0.5em]
N_q=\frac{\zeta(3)}{\pi^2}\,g_*^i T^3 \end{array}\right. 
,~g_*^i=51.25 \\[0.5em]
\textrm{After}&\left\{\begin{array}{l}\rho  =\frac{\pi^2}{30}\,g_*^f T^4\\[0.5em]
s  =\frac{2\pi^2}{45}\,g_*^f T^3\\[0.5em]
N_h=\frac{\zeta(3)}{\pi^2}\,g_*^f T^3 \end{array}\right. ,~g_*^f=17.25,~\label{eq:BeforeAfter}
}
where $\zeta(3)\simeq 1.202$, for one family of quarks and leptons, gluons, one set of pions, photons, and more neutrinos. Note that the ratio of entropy to number density is $s/N=\frac{2\pi^4}{45\zeta(3)} \simeq 3.60174$.

Including the Hubble expansion, $\thb$ evolves according to
\dis{
 \ddot\thb+3H\dot\thb+ m_a^2(t)\sin\thb=0,\label{eq:thbevolution}
}   
where the angle $\thb$ is $a/f_a $ and $m_a(t)$ is the temperature dependent axion mass. At a cosmic time scale  $ m_a\sim 3H$, $\ddot{\thb}$ is negligible and Eq. (\ref{eq:thbevolution}) determines an angle $\thb_1$ which was known before as $T_1\simeq 1\,\gev$ \cite{Turner86,Bae08}. We will present new numbers below. At temperature $T_1$, the QCD phase is in the $\qg$-phase with the current quark masses, and hence the axion number determined is at the time when the single particle effect is dominant. In this region, the scattering effects in scattering experiments between single particles are expanded  in powers of $\Lambda_{\rm QCD}/|Q|$ where $Q^2$ is the momentum transfer in the high energy   scattering, where  $\Lambda_{\rm QCD}^{(3)}\simeq 332\,\mev$ for three light quarks \cite{QCD16}. On the other hand, after the QCD phase transition, it belongs to the many body phenomena where the quantity susceptibility $\chi$ is used. Fortunately, now there exist  numbers on  the susceptibility from lattice calculations at the level $\chi\simeq 76\,$MeV \cite{Boyarski16,ICTP16,Sharma16}.   
 
\begin{figure}[!t]
\includegraphics[width=0.35 \textwidth]{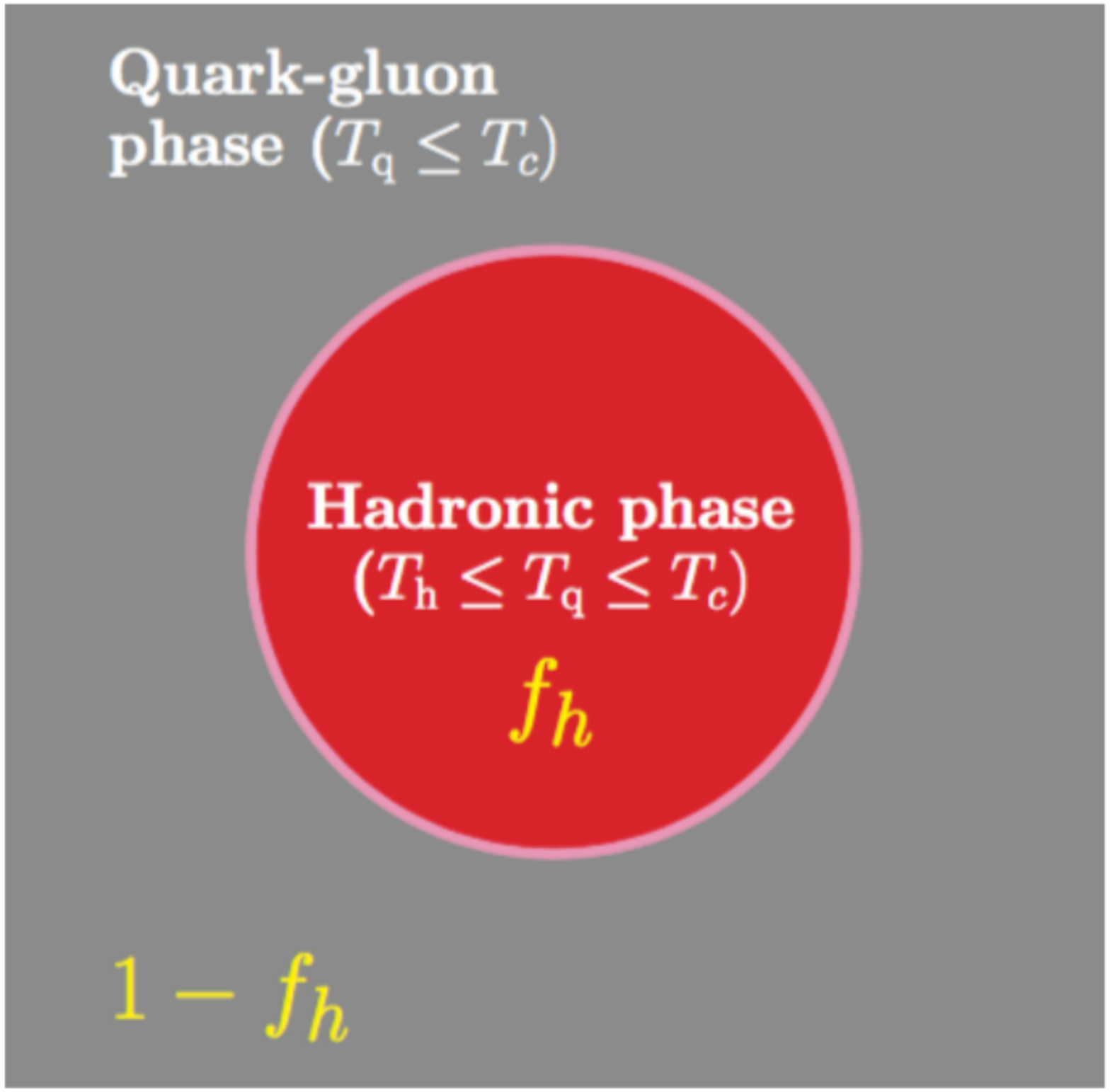} \hskip 1cm
\includegraphics[width=0.25 \linewidth]{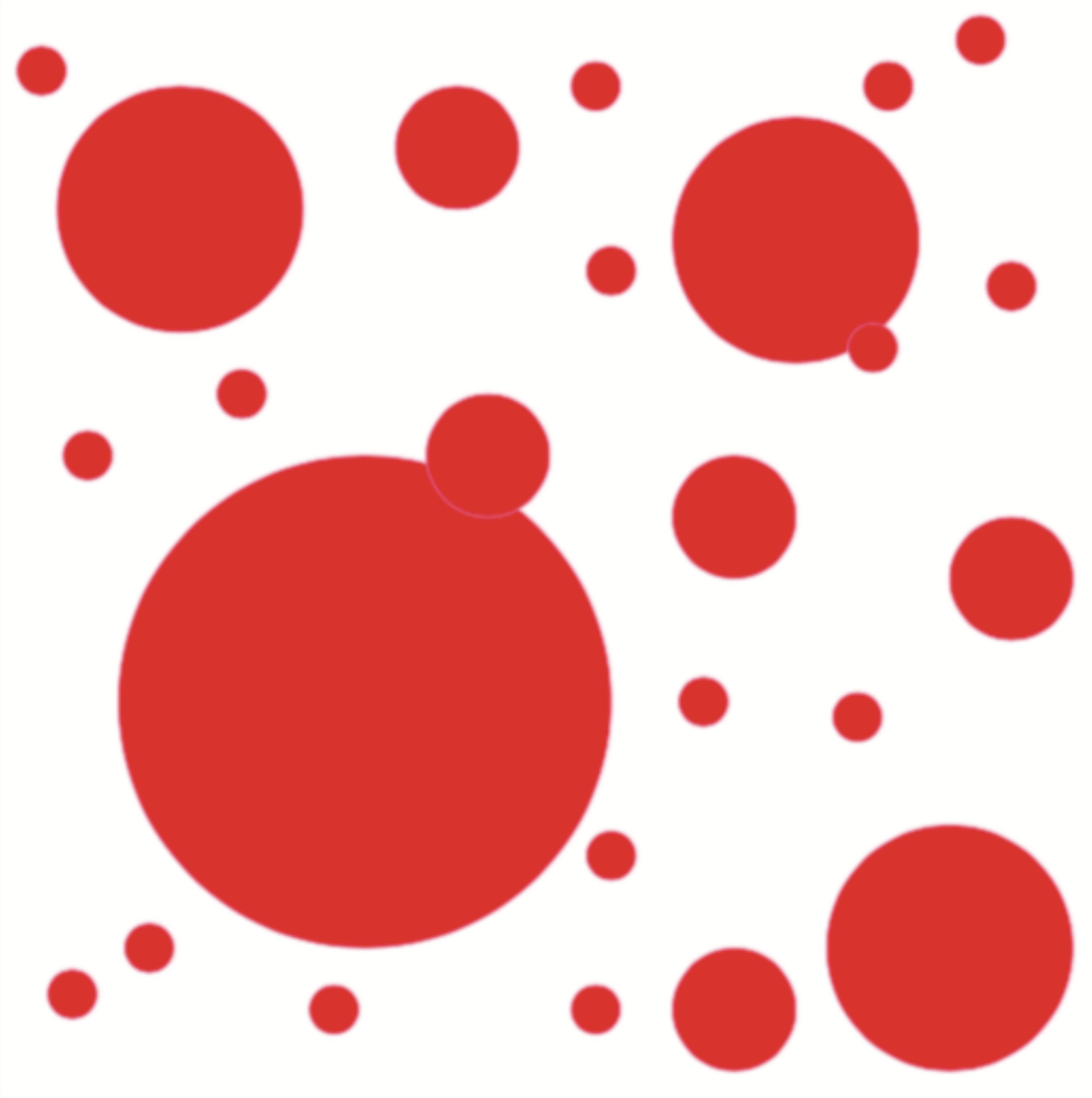} 
  \caption{Formation of hadronic bubbles at $T<T_c$. In the right figure, four scales of bubbles formed at four different time scales are shown.} \label{fig:hBubble} 
\end{figure}

In the left panel of Fig. 2,  formation of a typical size $h$-phase bubble  in the $\qg$-phase region at temperature $T_q$ is shown. Here, the $h$-phase fraction is $f_h$ and  the $\qg$-phase fraction is $(1-f_h)$. For the temperature--time relation in the evolving Universe, we use the spin degrees given in Eq. (\ref{eq:BeforeAfter}) and for comparing the number densities in the $\qg$- and $h$-phases  we use $g_*^{\rm q}=37$ and $g_*^{\rm h}=1$.\footnote{For one set of pions, we can use  $g_*^{\rm h}=3$ but counting just the number of pions, disregarding the charge differences,  it is more convenient to use $g_*^{\rm h}=1$ and move the factor 3 in the other equations.} 
In the right panel of Fig. \ref{fig:hBubble}, four scales of bubbles are illustrated at four different time scales.  
 
Here,  we use the field theoretic idea of bubble formation \cite{Okun74,Coleman77,CallCol77}, even though not using their first order form, but will solve a phenomenological differential equation of 
$f_h$,  the fraction of $h$-phase bubbles in the Universe,  introducing the rate $\alpha$. 
It is consistent with the observation of cross-over phase transition \cite{LosAlamos14}. 
 
 Depending on the independent thermodynamic variables, energies have different names,
\begin{eqnarray}
&dU=dQ-PdV+\mu dN,\label{eq:U}\\
&dA=-SdT-PdV+\mu dN,\label{eq:A}\\
&dG=-SdT+VdP+\mu dN,\label{eq:G}
\end{eqnarray}
where $U,A,$ and $G$ are  internal energy, free energy, and Gibbs free energy, respectively, and $\mu$ is  the chemical potential  (the energy needed to add one particle to a thermally 
and mechanically isolated system \cite{HuangStat}), and $N$ is the number density. Different energies are used for different physics:   $dU=0$ for the first law of thermodynamics, $dA=0$   in the expanding Universe, and $dG=0$ in the first order phase transition.\footnote{In our cross over phase transition, it starts like the first order and   ends like the second order.}  In  the beginning of the phase transition, we use $dG=0$,
\dis{
&\textrm{Phase change between \qg-   and  h-phases}:\delta G=-g_q\delta \mu_q+g_h\delta \mu_h=0,\label{eq:ConsGibbs}
}
where both signs of $\delta\mu_q$ and $\delta\mu_h$ are taken to be positive for one particle increment, and $g_{q,h}$ are the Gibbs free energies in the $\qg$ and $h$ phases.
During the QCD phase transition, therefore, the temperatures of quarks, gluons, and pions remain the same. Thermalization of hadrons with leptons changes temperature a bit and we use the resultant cosmic temperature as $T$.  To apply Eq. (\ref{eq:ConsGibbs}), we must know the pressure in both phases. Pressure in the $\qg$ phase is given in Eq. (\ref{eq:BeforeAfter}) as `Before'. But, we cannot use `After' of  Eq. (\ref{eq:BeforeAfter})  because it corresponds to particles well separated while our case of strong interaction is for the overlapping waves. We first calculate pressure for the overlapping waves and then use the conservation of Gibbs free energy to estimate the transition rate $\alpha$. These are calculated step by step:
\begin{itemize}
\item First calculate the average pion energy which does not depend on the overlapping nature of waves. At $T_c$, Eq. (\ref{eq:ConsGibbs}) gives  
\dis{
N_q\mu_q =N_\pi\mu_\pi\to
37\cdot \frac{\zeta(3)}{\pi^2}T_c^{3}\cdot T_c=N_\pi\cdot E_\pi, 
}
where $E_\pi$ is the pion energy. Thus, we obtain the number density $N_\pi$ for the average pion energy $\langle E_\pi\rangle$, 
\dis{
N_\pi(T_c)=\left(\frac{37}{3.60174}\, \frac{T_c}{\langle E_\pi\rangle}\right)T_c^{3}\simeq 3.8033\, T_c^3,
}
where the average energy at $T_c$ for a relativistic boson   is used \cite{KolbBk}
\dis{
\langle E_\pi\rangle =\frac{\pi^4}{30\zeta(3)}T_c\simeq 2.701\,T_c.
}

\item To calculate pressure, consider a perpendicular wall on which force is acted. Momentum change by the wall perpendicular to $x$-axis is $2Ev_x$, and the resulting force is $2Ev_x^2/\Delta  x$; thus the force on the unit area is $2Ev_x^2/\Delta x/{\rm area}=2Ev_x^2 \cdot({\rm number~density})\to \frac23 E(v_x^2+v_y^2+v_z^2 )\cdot({\rm number~density})$. For strongly correlated pions, therefore, we  obtain \cite{HuangStatRel}
\dis{
P_h(T)&= 3.8033\, T^3\times  \frac{2 }{3 }
\int_{m_\pi}^{E_{\rm cutoff}}  \frac{E}{e^{-\beta \mu}e^{-\beta(E-m_\pi)}-1} 
\frac{4\pi \sqrt{E^2-m_\pi^2}dE}{(2\pi)^3\,E^2},\\
&=- 0.1285\, m_\pi T^3
\int_1^{\sqrt{ 1+(T^2/m_\pi^2)}} \frac{\sqrt{x^2-1}}{1-e^{\frac{m_\pi}{T}(1-2x)} }
\frac{dx}{x},\label{eq:P}
}
where we  considered wavelengths up to the Compton wave length of $\pi$, $\lambda_{\rm min}\le \lambda\le  1/m_\pi$. Beyond $\lambda_{\rm min}$, it is better to consider quarks and gluons rather than pions. For $ \lambda\ge  1/m_\pi$, pions are considered to be individual particles. At $T$, the maximum pion energy is considered to be $E_{\rm max}=\sqrt{m_\pi^2+T^2}$. Thus, we used $E_{\rm cutoff}=\sqrt{m_\pi^2+T^2}$,
 for which we obtain the solid curve of Fig. \ref{fig:hPalpha}\,(a).  

\begin{figure}[!h]
  \includegraphics[width=7cm,height=4.5cm]{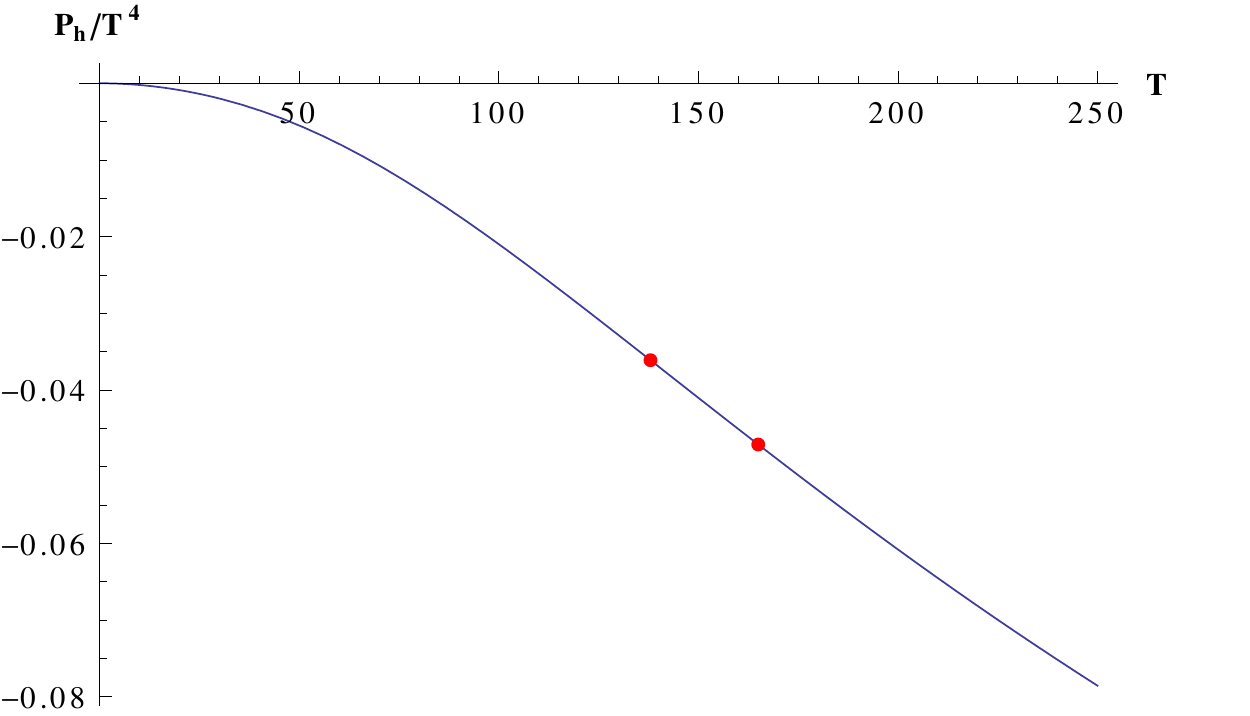}
   \includegraphics[width=7cm,height=4.5cm]{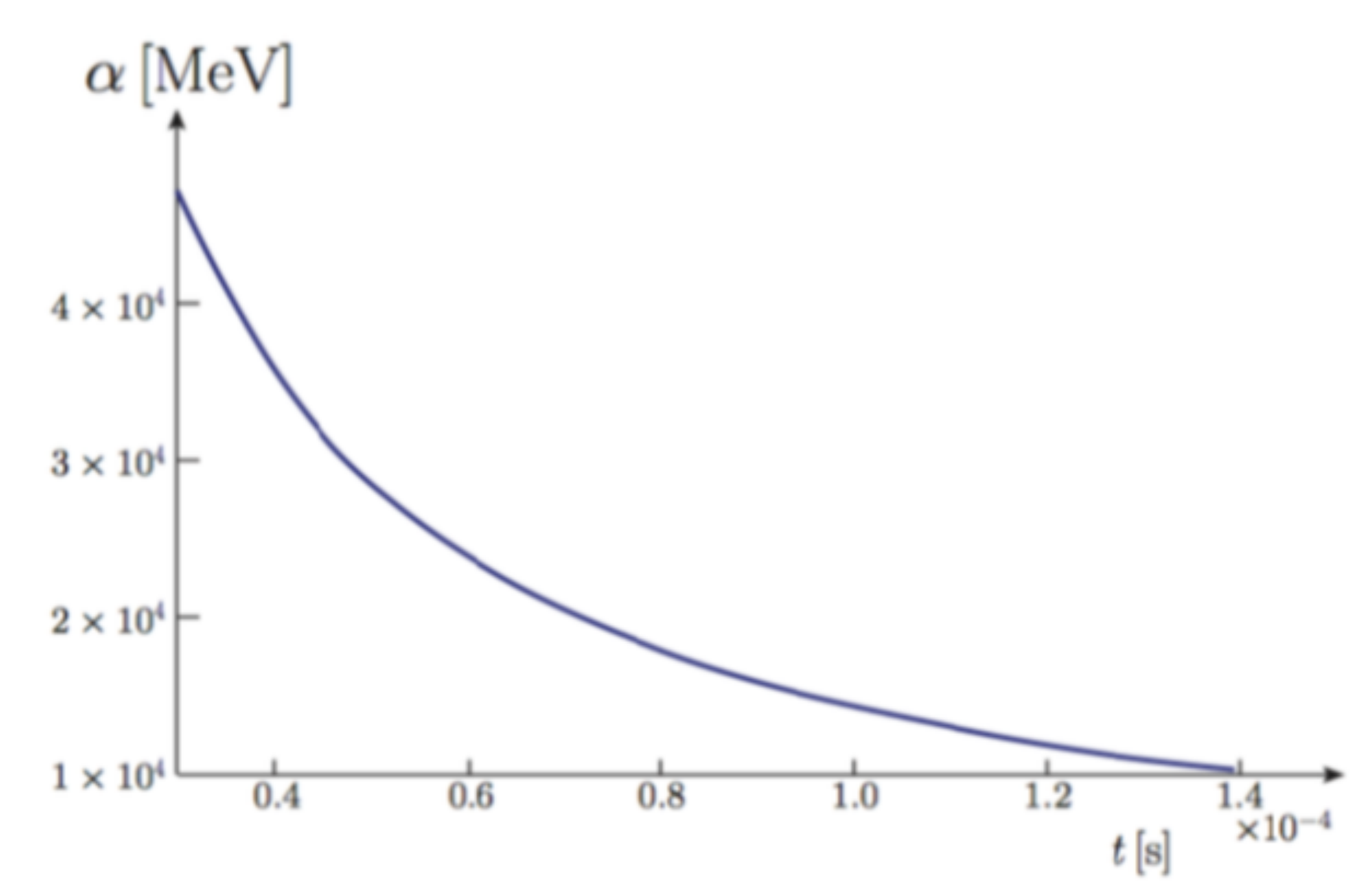}\\
   \centerline{(a)\hskip 5.5cm (b)}
\caption{(a) $P_h$ for  the overlapping waves versus  $T$, and (b)  $\alpha$ as a function of  $t$ for $C=1.5$.}\label{fig:hPalpha}
\end{figure}
 
\item After the $h$-bubbles are formed, the temperature (obtained by collisions)  of the inside $h$-phase drops faster than that of the outside $q$-phase because $g_*({\rm inside})<g_*({\rm outside})$ and pions are massive. This is illustrated as the temperature inequality in Fig. \ref{fig:hBubble}.  The expansion of bubbles  will be approximated by phenomenological parameters. 

\item The expanding Universe is the case of different pressures. So, we do not use $dG=0$ but consider $dA=0$,
\dis{
(-SdT-PdV+\mu dN)_q+(-SdT-PdV+\mu dN)_h=0.
}
Using $dV_q=-dV_h$,
\dis{
(P_h-P_q)dV_h =(S_q -S_h)dT +\mu_h dN_h-\mu_q dN_q= (S_q -S_h)dT.
}
\dis{
\frac{1}{V}\frac{dV_h}{dt} =\frac{(S_q -S_h)}{(P_h-P_q)}
\frac{dT}{dt} .\label{eq:alpha0}
}
Since $dT/dt$ is negative, the right-hand side of Eq. (\ref{eq:alpha0}) is always positive and the fraction of $h$-phase  increases. 
 Note that it is close to the  change rate of fraction of $h$ phases in the whole Universe.
The RHS of Eq. (\ref{eq:alpha0}) is defined as the rate  of formation of the $h$-phase ball of radius $R_i$ of typical pion size,  $\alpha(T)$,
\dis{
\alpha(T) \approx \frac{- 37\pi^2
}{ 45(P_h-P_q)}\frac{T^6}{\mev} ,~  {\rm with}~T^2t_{\rm [s]}\simeq \mev,\label{eq:alphaT}
}
in which we used $dT/dt = -\frac12 T/t$ in the radiation dominated (RD) Universe.  In Fig. \ref{fig:hPalpha}\,(b), $\alpha(T)$ is shown.
\end{itemize}
 
\begin{figure}[!h]
 \includegraphics[width=0.55\textwidth]{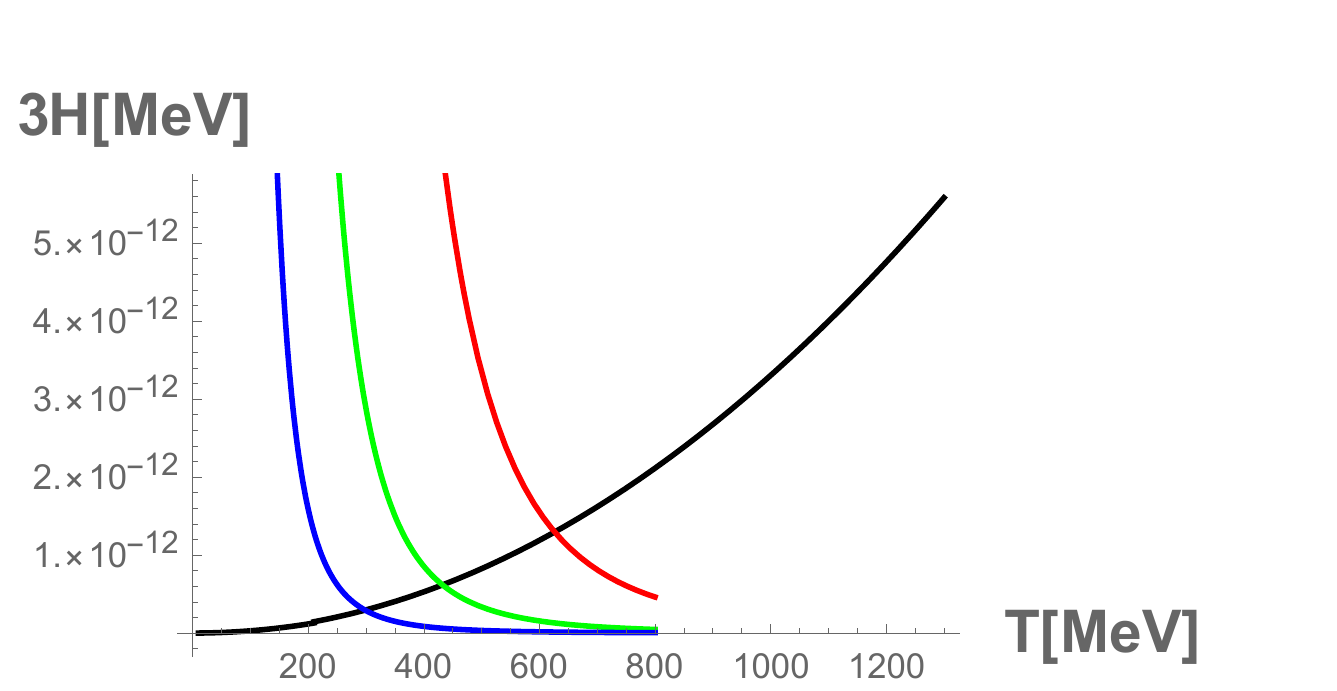}
 \includegraphics[width=0.4\textwidth]{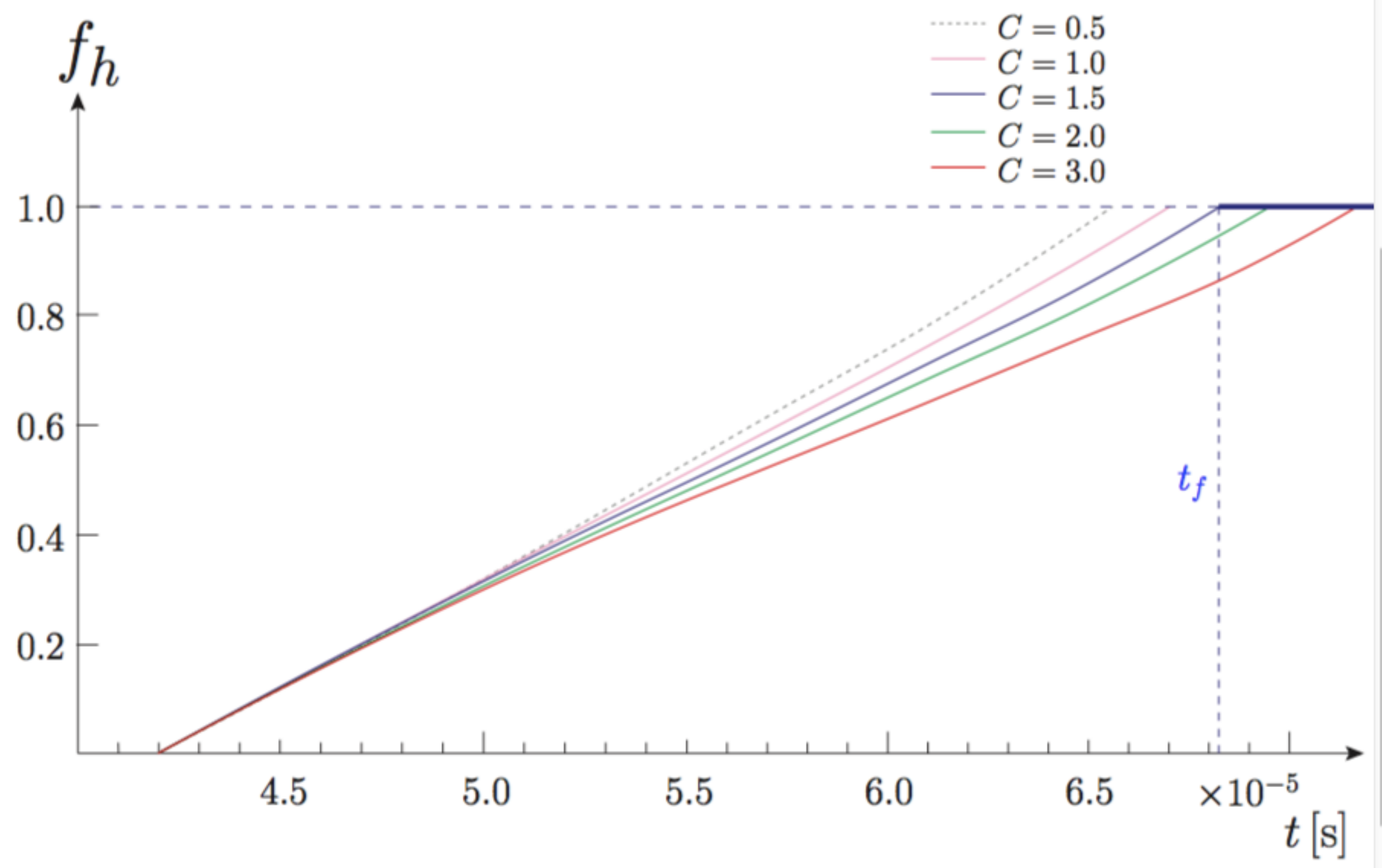}
\caption{(a) $m_a(T=0)$ and $3H$ vs. $T$, and (b) $f_h$ as a function of $t$. In (a), different $T_1$'s are given for different  $m_a(T=0)$'s at $\thb_1=1$: $T_1= 0.298\,\gev,0.432\,\gev, 0.626\,\gev$ for $m_a(0)= 10^{-5}\,\eVV$(blue), $10^{-4}\,\eVV$(green), and $ 10^{-3}\,\eVV$(red), respectively. }\label{fig:T1C}
\end{figure}

For the critical temperature, we use  $T_c=165\pm 5\mev$ \cite{Boyarski16,ICTP16}. At $T_c$, two phases co-exist. So, $\chi$ in the  $h$-phase  at $T_c$  is equated to the $\chi$   in the  $\qg$-phase at $T_c$. Knowing  the $\chi$ value in the  $\qg$-phase at $T_c$, we extend it to the GeV region via the instanton effect with temperature dependence. In the $\qg$-phase above the $\rho$ meson mass scale, we use the temperature dependence  $ T^{-8.16}$ \cite{Pisarski81}. We use this power in the region where quarks and gluons are manifest, \ie above the $\rho$ meson mass scale. Below the $\rho$ meson mass scale and above $T_c$, the constituent quark mass around 300 MeV are present.  We smoothly connect the temperature dependence in this region, between the $\rho$ meson mass scale and $T_c$, by an interim power $  T^{-4.21}$.\footnote{It was noted in Ref. \cite{Sharma16} that  the temperature dependence of topological susceptibility is very different from dilute instanton gas approximation and mimicks it  from $T>250$ MeV. Our power $4.21$ is partly in accord with this observation.
}
 Then, determine $\thb_1$ by the condition $m_a(T_1)=3H(T_1)$. The temperture $T_1$ is determined as shown in Fig. \ref{fig:T1C}\,(a). $T_1$ depends on the zero temperature axion mass $m_a(0)$. If $m_a(0)$ is smaller than $4\times 10^{-3}\eVV$, $T_1$ is below the $\rho$ mass scale  $770\,\mev$.

 Some time after a bubble is formed, its radius would expand  with the light velocity, and its volume would increase with the rate $3/R(t)$ where $R(t)=t+R_i$. Here, $R_i$ is the initial radius of the nucleated $h$-phase ball,  $R_i=R(0)$.
Therefore, including the Hubble expansion, the differential equation for fraction $f_h$ of $h$-phase is approximated by the following  differential equation, summarizing effectively the above items by two parameters,  $\alpha(T) $ and $C$,
\dis{
\frac{df_h}{dt }=\alpha (1-f_h)+\frac{3}{ {[1+C f_h(1-f_h) ]}(t+R_i)}f_h,  \label{eq:diffequa}
}
where  $\alpha(T) $ of Eq. (\ref{eq:alphaT}) is for the formation of the $h$-phase ball of radius $R_i$.
 The initial condition is $f_h(t=0)=0$. In Eq. (\ref{eq:diffequa}), $3/(t+R_i)$ takes into account  expansion of a bubble(s), starting from radius $R_i$. As time goes on, some bubbles coalesce and the overlapped part should not be considered for expanding. This overlapping part diminishes as $f_h$ approaches 1 since there is not much $(1-f_h)$ from which new bubbles would form. Near $f_h=0$ also, there is no coalesce effect since the balls have not expanded yet. Equation (\ref{eq:diffequa}) introduces a phenomenological parameter $C(>0)$ to take into account these coalesce  effects. In Fig. \ref{fig:T1C}\,(b), we show   $f_h(t)$   for several different values of $C$.
 The  $C$ dependence controls the value $T_f$.  For example,   $C=0.5, 1, 1.5, 2,$ and 3 give  
 $T_f=0.653T_c, 0.647T_c, 0.641T_c, 0.634T_c,$ and $ 0.623T_c$, 
 respectively.\footnote{In Ref. \cite{DeGrand84}, it was argued that $T_f>0.6 T_c$.}  The $C$ dependence is not very dramatic, and we use   $C=1.5$ in Fig. \ref{fig:rhoa} for which  the Hubble radius is increased by a factor of $\simeq 2.4$ during this QCD phase transition.  

As the Universe expands, the QCD phase transition starts at $T_c$, and ends when  $f_h=1$ is reached, whose time scale is denoted as $t_f$ (at temperature $T_f$). It is illustrated as the dashed curve for $C=1.5$ in the lavender box in Fig. \ref{fig:rhoa}. Then the phase transition is complete, after which   the Universe goes into the RD  in the  $h$-phase.    
  
If $x$ fraction  of the current CDM energy density is made of ``invisible'' axions, 
the axion energy density is $x$ times the current critical energy density,
 \ie numerically $x\times 0.9935\times 10^{-35}\mev^4$. In the expanding Universe, from this value at $t_{\rm now}$  the  ``invisible'' axion energy density at $t_f$ is estimated as $\rho_a(t_f)$ shown in Fig. \ref{fig:rhoa}. These two values are related as     $\rho_a({\rm now})\simeq \rho_a(t_f)\cdot( \thb_{\rm now}/\thb_f)^2$.  Since   $\rho_a(t_f)$ calculated through the QCD phase transition earlier in this section   is O$(10^7 \mev^4\thb_f^2)$ from the scale in Fig. \ref{fig:rhoa}, $(\thb_{\rm now}/\thb_f)^2$ must be of order $10^{-42}x$. Thus, for the ``invisible'' axion to become CDM, $|\thb_{\rm now}/\thb_f|$ must be of order $10^{-21}\sqrt{x}$.  
  
\section{$\thb$ evolution in the bottle neck period and more}\label{sec:Bottleneck}
\begin{figure}[!t]
 \includegraphics[width=10cm,height=7cm]{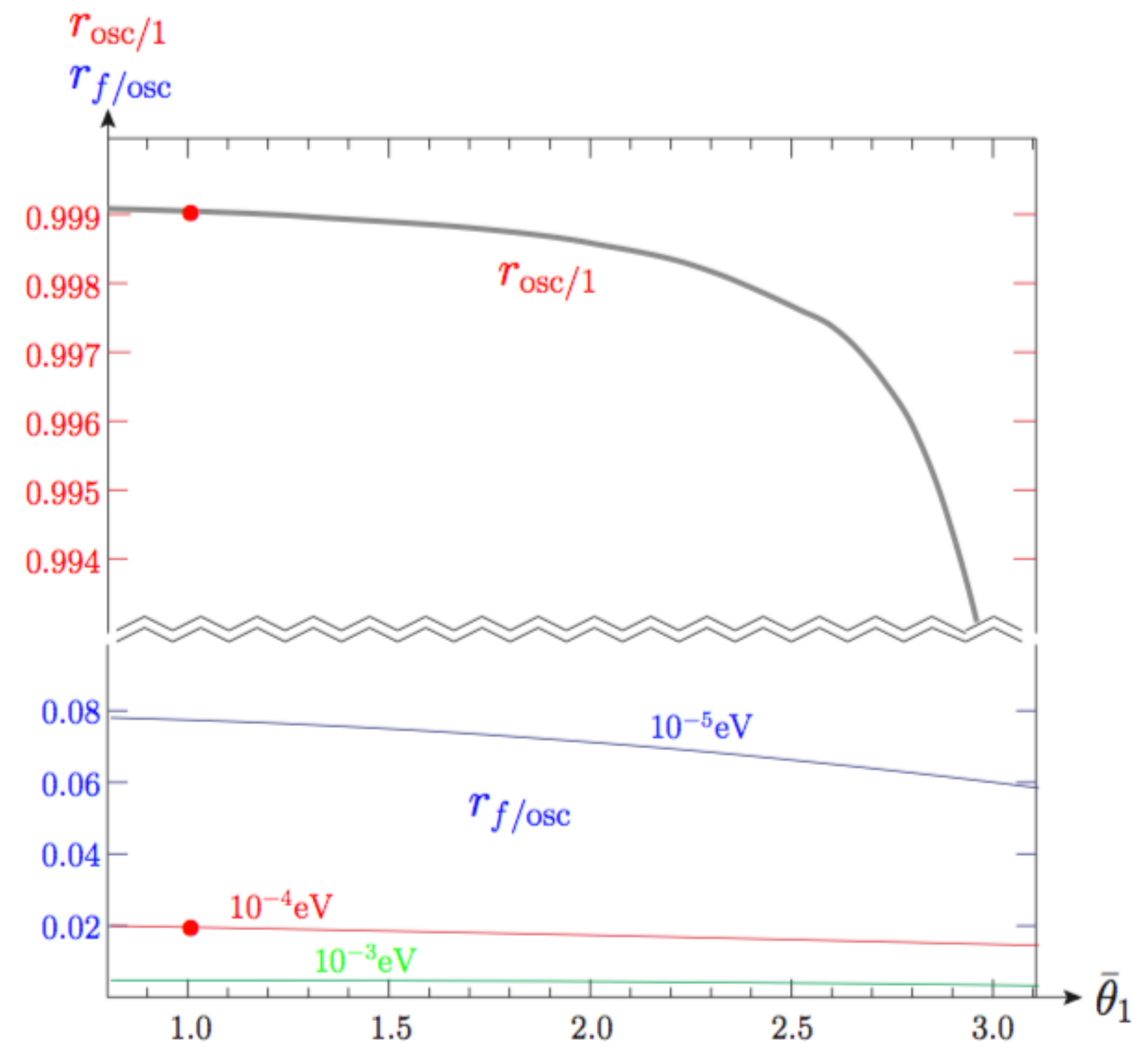}
\caption{The ratios  $r_{\rm osc/1}\equiv \bar{\theta}_{\rm osc}/\bar{\theta}_1$ and $r_{f /\rm osc} \equiv \bar{\theta}_f/\bar{\theta}_{\rm osc}$  as  functions of $\bar{\theta}_1$ for three $m_a(0) = 10^{-3}\eVV (\rm green), 10^{-4}\eVV (\rm red),$ $10^{-5}\eVV (\rm blue)$. In the upper figure, these curves are almost overlapping  shown as gray.  $t_{\rm osc}$ is the time of the 1st oscillation after which the harmonic motion is a good description. Different $T_1$'s  are used for different $m_a(0)$, as presented in Fig. \ref{fig:T1C}.   }\label{fig:ratioOsc}
\end{figure}

We determined $T_1$ by the condition $m_a(T_1)=3H(T_1)$. Then, from $T_1$  to $T_{\rm osc}$,  use the evolution equation of $\thb$,
\dis{
 \ddot\thb +3H\dot{\thb} + m_a^2(t)\sin \thb=0,\label{eq:diffeqtau}
}
where dot denotes the derivative with respect to $t$.    After  $t_{\rm osc}$, the harmonic oscillation is an excellent description of the oscillation \cite{Bae08,KimKimNam18}. 
Figure  \ref{fig:ratioOsc} shows the factors  $r_{\rm osc/1}$ (the upper figure) the ratio of $\thb$'s at the time $t_{\rm osc}$ (the commencement  time of the 1st oscillation after the bottle neck period)  and at $t_1$, and   $r_{f/ \rm osc}$ (the lower figure)  the ratio at $t_f$ and at  $t_{\rm osc}$. Three curves are for three axion masses,  $m_a=10^{-3}\eVV,10^{-4}\eVV$, and $10^{-5}\eVV$.   $r_{\rm osc/1}$ does not have a strong dependence on the axion mass, but $r_{f/\rm  osc}$ has the axion mass dependence as shown in the lower part in Fig.  \ref{fig:ratioOsc}. For $m_a=10^{-4}\eVV $, Fig. \ref{fig:ratioOsc} shows $r_{\rm osc/1}=0.99871, 0.99871\footnote{ This  number  becomes 0.99437 if 332 MeV is used instead of  $m_\rho$ for the cusp position in the inset of Fig. \ref{fig:rhoa}.},   0.97407$, and $r_{f/ \rm osc}=2.005\times 10^{-2},2.005\times 10^{-2},$ and $1.5346\times 10^{-2}$, respectively, for $\thb_1=0.5, 1, 0.99\pi$. Bullets correspond to $\thb=1$ for $m_a=10^{-4}\eVV $, in which case the product  is $r_{f/1}(\thb_1=1)\equiv r_{\rm osc/1}(1)\cdot r_{f/\rm /osc}(1)\simeq 2.002\times 10^{-2}.$  
Figure \ref{fig:rhoa} takes these effects into account, specifically for $m_a=10^{-4}\eVV $ while the Universe was evolving.  

From   Fig.  \ref{fig:ratioOsc},  we obtain an approximate formula for $r_{f/1}$ in 
the range $m_a=[10^{-3}\eVV,10^{-5}\eVV]$,
\begin{equation}
r_{f/1}\simeq 0.02\left(\frac{m_a}{10^{-4}\eVV}\right)^{-0.591\pm 0.008},\label{eq:ratio}
\end{equation}
where the error bars are given from possible ranges of curves in Fig.  \ref{fig:ratioOsc}.
The power 0.591 can be compared to 0.184 of Ref.  \cite{Bae08}
and $\frac16$ of Ref. \cite{Sikivie08}. Our large value is due to our method of obtaining different $T_1$'s for different axion masses shown in Fig. \ref{fig:T1C}\,(a), in contrast to using a unique value for $T_1$ \cite{Bae08,Sikivie08}.
We stress again that the overall coefficient 0.02 is for the case of $m_a =10^{-4}\eVV$ and  the power in Eq. (\ref{eq:ratio}) corrects for the mass difference effect in the range we consider. If one gives $\bar{\theta}_1$,  $\bar{\theta}_f$ is calculated by  Eq. (\ref{eq:ratio}), and   $\rho_a({\rm now})\simeq \rho_a(t_f)\cdot( \thb_{\rm now}/\thb_f)^2$.  $\rho_a(t_f)$ is read in Fig. \ref{fig:rhoa} and $\thb_{\rm now}/\thb_f$ is reliably calculated recently in Ref. \cite{KimKimNam18} .

\section{Conclusion} 

We studied the QCD phase transition, satisfying two conditions: coexistence of $\qg$- and $h$-phases with the same Gibbs free energy below $T_c$  and the expansion of the $h$-phase bubbles afterwards by a phenomenological differential equation, Eq. (\ref{eq:diffequa}). These allow a narrow temperature range for forming $h$-phase bubbles at $T\le 165\,\mev$. 
   Within this narrow range, we obtain a temperature dependent bubble formation rate   $\alpha(T)$.  Using this $\alpha(T)$, we obtain the completion temperature of the QCD phase transition,   $T_f\simeq 126\,\mev$, corresponding to $t_f\approx  63\,\mu{\rm s}$.  This provides a key knowledge on the energy density of ``invisible''  QCD axion at  $t_f$ which allows us to estimate the current CDM density of   ``invisible''  QCD axion.

 \section*{Acknowledgments}
 J.E.K. thanks Deog Ki Hong, Duchul Kim, and A. Ringwald for helpful communications.
This work is supported in part by the  NRF grant funded by the Korean Government (MEST)
(NRF-2015R1D1A1A01058449). J.E.K. is supported also in part by IBS-R017-D1-2018-a00, and S. K.  supported in part also by NRF-2015aR1D1A1A09059301. 
  


\begin{thebibliography}{99}

\def\apj#1#2#3{{Astrophys.\ J.}\ {\bf #1} (#3) #2}
\def\apjs#1#2#3{{Astrophys.\ J. Supp.}\ {\bf #1} (#3) #2}
\def\apph#1#2#3{{Astropart. Phys.}\ {\bf #1} (#3) #2}
\def\anj#1#2#3{{Astronom.\ J.}\ {\bf #1} (#3) #2}
\def\anap#1#2#3{{Astronom.\ Astrophys.}\ {\bf #1} (#3) #2}
\def\mnras#1#2#3{{Mon.\ Not.\ R.\ Astron.\ Soc.}\ {\bf #1} (#3) #2}
\def\anrnp#1#2#3{{Annu.\ Rev.\ Nucl.\ Part.\ Sci.}\ {\bf #1} (#3) #2}
\def\ijmpa#1#2#3{{Int.\ J.\ Mod.\ Phys. A}\ {\bf #1} (#3) #2}
\def\mpl#1#2#3{{Mod.\ Phys.\ Lett.\ A}\ {\bf #1} (#3) #2 }
\def\nat#1#2#3{{Nature}\ {\bf #1} (#3) #2}
\def\npb#1#2#3{{Nucl.\ Phys.\ B}\ {\bf #1} (#3) #2}
\def\plb#1#2#3{{Phys.\ Lett.\ B}\ {\bf #1} (#3) #2}
\def\prd#1#2#3{{Phys.\ Rev.\ D}\ {\bf #1} (#3) #2}
\def\prx#1#2#3{{Phys.\ Rev.\ X}\ {\bf #1} (#3) #2}
\def\pr#1#2#3{{Phys.\ Rev.}\ {\bf #1} (#3) #2}
\def\prl#1#2#3{{Phys.\ Rev.\ Lett.}\ {\bf #1} (#3) #2}
\def\prp#1#2#3{{Phys.\ Rep.}\ {\bf #1} (#3) #2}
\def\sjnp#1#2#3{{Sov.\ J.\ Nucl.\ Phys.}\ {\bf #1} (#3) #2}
\def\zp#1#2#3{{Z.\ Phys.}\ {\bf #1} (#3) #2}
\def\jhep#1#2#3{{JHEP}\ {\bf #1} (#3) #2}
\def\epjc#1#2#3{{Euro. Phys. J. C}\ {\bf #1} (#3) #2}
\def\rmp#1#2#3{{Rev. Mod. Phys.}\ {\bf #1} (#3) #2}
\def\sci#1#2#3{{Science}\ {\bf #1} (#3) #2}
\def\prth#1#2#3{{Prog. Theor. Phys.}\ {\bf #1} (#3) #2}
\def\njp#1#2#3{{New J. Phys.}\ {\bf #1} (#3) #2}
\def\jkps#1#2#3{{J. Korean  Phys. Soc.}\ {\bf #1} (#3) #2}
\def\jetpl#1#2#3{{JETP Lett.}\ {\bf #1} (#3) #2}
\def\jcap#1#2#3{{JCAP}\ {\bf #1} (#3) #2}
  \def\cmp#1#2#3{{Comm. Math. Phys.}\ {\bf #1} (#3) #2}
\def\frp#1#2#3{{Front. Phys.}\ {\bf #1} (#3) #2}
\def\mpla#1#2#3{{Mod. Phys. Lett. A}\,{\bf #1} (#3) #2}
\def\err#1#2#3{Erratum: {\it ibid.} {\bf #1},  #2  (#3)}

\bibitem{KSVZ1} J. E. Kim,   \emph{Weak interaction singlet and strong CP invariance}, \prl{43}{103}{1979} [doi: 10.1103/PhysRevLett.43.103].

\bibitem{KSVZ2}  M. A. Shifman, V. I. Vainstein, V. I. Zakharov,   \emph{Can confinement ensure natural CP invariance of strong interactions ?},
\npb{166}{4933}{1980} [doi: 10.1016/0550-3213(80)90209-6].

\bibitem{DFSZ}
M. Dine, W. Fischler, and M. Srednicki,   \emph{A simple solution to the strong CP problem with a harmless axion}, \plb{104}{199}{1981} [doi: 10.1016/0370-2693(81)90590-6].

\bibitem{DFSZ2}
A. R. Zhitnitsky,    \emph{On possible suppression of the axion hadron interactions}, {Sov. J. Nucl. Phys.} {\bf 31} (1980) 260.

  \bibitem{PQ77} R. D. Peccei and H. R. Quinn, \emph{CP conservation in the presence of instantons}, \prl{38}{1440}{1977} [doi:10.1103/PhysRevLett.38.1440].
   
\bibitem{Preskill83} J. Preskill, M. B. Wise, and F. Wilczek,  \emph{Cosmology of the invisible axion}, \plb{120}{127}{1983}. [doi: 10.1016/0370-2693(83)90637-8].

\bibitem{Abbott83}
L. F. Abbott and P. Sikivie,    \emph{A cosmological bound on the invisible axion}, \plb{120}{1983}{133} [doi: 10.1016/0370-2693(83)90638-X].

\bibitem{Dine83}
M. Dine and W. Fischler,    \emph{The not so harmless axion}, \plb{120}{137}{1983} [doi: 10.1016/0370-2693(83)90639-1].
 
 \bibitem{Weinberg78} S. Weinberg, \emph{A new light boson?}, \prl{40}{223}{1978} [doi:10.1103/PhysRevLett.40.223].

 \bibitem{Wilczek78} F. Wilczek, \emph{Problem of strong p and t invariance in the presence of instantons}, \prl{40}{279}{1978} [doi:10.1103/PhysRevLett.40.279].

\bibitem{Boyarski16}  Sz. Borsanyi {\it et al.},  \emph{Calculation of the axion mass based on high-temperature lattice quantum chromodynamics}, 
\nat{539}{69}{2016} [arXiv:1606.07494 [hep-lat]].
 
 \bibitem{KimSemTsu14} J. E. Kim, Y. K. Semertzidis, and S. Tsujikawa, \emph{Bosonic coherent motions in the Universe}, \frp{2}{60}{2014}  [arXiv:1409.2497 [hep-ph]].
  
\bibitem{Sikivie17} P. Sikivie, \emph{Gravitational self-interactions of a degenerate quantum scalar field}, Talk presented at 13th Patras Axion-WIMP Workshop, Thessaloniki, Greece,   15--19 May 2017. 

\bibitem{KimRMP10} J. E. Kim and G. Carosi,
\emph{Axions and the strong CP problem}, \rmp{82}{557}{2010} [arXiv:0807.3125 [hep-ph]].
.

 \bibitem{ADMX18} B. M. Brubaker  \etal\,(ADMX Collaboration), \emph{First results from a microwave cavity axion search at  24 $\mu\eVV$}, \prl{118}{061302}{2018} [arXiv:1610.02580 [astro-ph.CO]].

\bibitem{Turner86} M. S. Turner,  \emph{Cosmic and local mass density of invisible axions}, \prd{33}{889}{1986} [doi:10.1103/PhysRevD.33.889].

\bibitem{Bae08}  K. J. Bae, J-H. Huh, and J. E. Kim,   \emph{Updating the axion cold dark matter energy density}, \jcap{09}{005}{2009} [arXiv:0806.0497 [hep-ph]].
 
\bibitem{KimKimNam18} J. E. Kim, S-J. Kim, and S. Nam, \emph{Axion energy density during the bottle neck period and $\thb$ ratio between early and late times}, arXiv:1803.03517 [hep-ph]


\bibitem{BardeenTye78} W. A. Bardeen and S-H. H. Tye, \emph{Current algebra applied to properties of the light Higgs boson},  \plb{74}{229}{1978} [doi:10.1016/0370-2693(78)90560-9].

\bibitem{Baluni78}
 V. Baluni, \emph{CP violating effects in QCD}, \prd{19}{2227}{1979} [doi:10.1103/PhysRevD.19.2227].

\bibitem{ICTP16} G. Grilli di Cortona, E. Hardy,  J. P. Vega, and G. Villadoro,  \emph{The QCD axion, precisely}, \jhep{1601}{034}{2016} [arXiv:
1511.02867 [hep-ph]].

\bibitem{Pisarski81} D. J. Gross, R. D. Pisarski, and L. G. Yaffe,  \emph{QCD and instantons at finite temperature}, \rmp{53}{43}{1981} [doi:10.1103/RevModPhys.53.43].

\bibitem{KimPRP87} J. E. Kim, \emph{Light pseudoscalars, particle physics and cosmology}, \prp{150}{1}{1987} [doi:10.1016/0370-1573(87)90017-2].

\bibitem{Manohar16}  A. Manohar and C. T. Sachrajda,    \emph{Quark mass}, in C. Patrignani et al. (Particle Data Group), Chin. Phys. C {\bf 40}, 100001 (2016).

\bibitem{QCD16} S. Bethke, G. Dissertori, and G. P. Salam, \emph{Quantum chromodynamics}, in 
C. Patrignani et al. (Particle Data Group), Chin. Phys. C {\bf 40}, 100001 (2016).

\bibitem{Sharma16} P. Petreczky, H-P. Schadler, and S. Sharma,  
\emph{The topological susceptibility in finite temperature QCD and axion cosmology}, \plb{762}{498}{2016} [arXiv:1606.03145 [hep-lat]]. 

\bibitem{Kogut88} J. B. Kogut and D. K. Sinclair, \emph{The thermodynamics of SU(3) lattice gauge theory with a light isodoublet of quarks}, \npb{245}{480}{1988} [doi: 10.1016/0550-3213(88)90531-7].

\bibitem{Forcrand07} P. de Forcrand and O. Philipsen, \emph{The chiral critical line of $N_f =2+1$ QCD at zero and non-zero baryon density}, \jhep{0701}{077}{2007} [arXiv:hep-lat/0607017].

\bibitem{AokiY06} Y. Aoki, G. Endrodi, Z. Fodor, S. D. Katz, and   K. K. Szabo,
\emph{The order of the quantum chromodynamics transition predicted by the standard model of particle physics}, \nat{443}{675}{2006} [arXiv: hep-lat/0611014]. 

 \bibitem{HuangStat}  K. Huang, \emph{Introduction to Statistical Physics} (Taylor \& Francis, London, 2001). 
 
\bibitem{LosAlamos14} T. Bhattacharya \etal, \emph{QCD phase transition with chiral quarks and physical quark masses}, \prl{113}{082001}{2014} [arXiv:1402.5175 [hep-lat]].

 \bibitem{Okun74} M. B. Voloshin, I. Yu. Kobzarev, and L. B. Okun, \emph{Bubbles in metastable vacuum}, Yad, Fiz. {\bf 20} (1974) 1229 [Sov. J. Nucl. Phys. {\bf 20} (1975) 644].
  
\bibitem{Coleman77} S. Coleman, \emph{The fate of the false vacuum. 1. Semiclassical theory}, \prd{15}{2929}{1977} [doi:10.1103/PhysRevD.15.2929] and \err{16}{1248}{1977}[doi:10.1103/PhysRevD.16.1248].

\bibitem{CallCol77} 
C. G. Callan and S. R. Coleman, \emph{
 The fate of the false vacuum. 2. First quantum corrections}, \prd{16}{1762}{1977} [doi:10.1103/PhysRevD.16.1762]. 
 
 \bibitem{KolbBk} E. W. Kolb and M. S. Turner, \emph{The Early 
 Universe} (Addison-Wesley, Red Wood City, CA, 1990).
 
 \bibitem{HuangStatRel} The relativistic form of Eq. (8.55) in  Ref. \cite{HuangStat}. 

\bibitem{DeGrand84} T. DeGrand and K. Kajantie, \emph{Supercooling, entropy production, and bubble kinetics in the quark-hadron phase transition in the early universe}, \plb{147}{273}{1984} [doi:10.1016/0370-2693(84)90115-1].

\bibitem{Sikivie08} P. Sikivie, \emph{Axion cosmology}, in Lect. Notes Phys. {\bf 741} (2008) 19-50 [astro-ph/0610440 
]. 
 
\end{thebibliography}
\end{document}